Label-free evaluation of myocardial infarct in surgically excised ventricular myocardium by Raman spectroscopy

Tsunehisa Yamamoto[1,2,¶], Takeo Minamikawa[1,3,4,¶], Yoshinori Harada[1], Yoshihisa Yamaoka[5], Hideo Tanaka[1], Hitoshi Yaku[2], and Tetsuro Takamatsu[1,6]

[1]Department of Pathology and Cell Regulation, Graduate School of Medical Science, Kyoto Prefectural University of Medicine, 465 Kajii-cho Hirokoji Kawaramachi, Kamigyo-ku, Kyoto 602-8566, Japan

[2]Department of Cardiovascular Surgery, Graduate School of Medical Science, Kyoto Prefectural University of Medicine, 465 Kajii-cho Hirokoji Kawaramachi, Kamigyo-ku, Kyoto 602-8566, Japan

[3]Department of Mechanical Science, Division of Science and Technology, Graduate School of Technology, Industrial and Social Sciences, Tokushima University, 2-1 Minami-Josanjima, Tokushima, Tokushima 770-8506, Japan

[4]PRESTO, Japan Science and Technology Agency (JST), 2-1 Minami-Josanjima, Tokushima, Tokushima 770-8506, Japan

[5]Department of Advanced Technology Fusion, Graduate School of Science and Engineering, Saga University, 1 Honjo, Saga, 840-8502, Japan

[6]Department of Medical Photonics, Kyoto Prefectural University of Medicine, 465 Kajii-cho Hirokoji Kawaramachi, Kamigyo-ku, Kyoto 602-8566, Japan

[¶]Co-first author (equal contribution)

Abstract

Understanding the viability of the ischemic myocardial tissue is a critical issue in determining the appropriate surgical procedure for patients with chronic heart failure after myocardial infarction (MI). Conventional MI evaluation methods are; however, preoperatively performed and/or give an indirect information of myocardial viability such as shape, color, and blood flow. In this study, we realize the evaluation of MI in patients undergoing cardiac surgery by Raman spectroscopy under label-free conditions, which is based on intrinsic molecular constituents related to myocardial viability. We identify key signatures of Raman spectra for the evaluation of myocardial viability by evaluating the infarct border zone myocardium that were excised from five patients under surgical ventricular restoration. We also obtain a prediction model to differentiate the infarcted myocardium from the non-infarcted myocardium by applying partial least squares regression-discriminant analysis (PLS-DA) to the Raman spectra. Our prediction model enables identification of the infarcted tissues and the non-infarcted tissues with sensitivities of 99.98% and 99.92%, respectively. Furthermore, the prediction model of the Raman images of the infarct border zone enabled us to visualize boundaries between these distinct regions. Our novel application of Raman spectroscopy to the human heart would be a useful means for the detection of myocardial viability during surgery.

**Introduction**

Understanding the viability of the ischemic myocardium is a critical issue for surgical treatment options for the failing heart after myocardial infarction (MI). Various functional evaluations of the heart are conducted preoperatively by, e.g., computed tomography, magnetic resonance imaging, radioisotope imaging, and echocardiography,[1-4] to determine whether the ischemic myocardium has a potential to recover contractile functions after reestablishment of the coronary circulation.[5,6] However, these diagnostic modalities are inadequate to detect precise, regional myocardial dysfunction in surgical situations, because of their relatively poor regional correlation with the real heart under direct vision. During heart surgery, viability of the myocardium is evaluated merely from the appearance of the heart *in situ* by reference to the indirect preoperative assessments. Thus, deeply desired is a useful means for intraoperative evaluation of the myocardial viability under direct vision to obtain better outcome of cardiac surgery.

Raman spectroscopy is expected to be a valuable analytical tool in the biomedical research field, allowing label-free, functional imaging of biological samples via molecular vibrations with no need for fixation or staining.[7-13] For over two decades, this modality has been applied experimentally to a variety of human tissues, e.g., the brain,[14] coronary artery,[15] skin,[16] breast,[17,18] and peripheral nerves.[19-21] In heart tissue, however, the great challenge for clinical application of the conventional Raman spectroscopy is because of the weak signals. We previously reported Raman spectroscopic analysis of old myocardial infarct of rats created by coronary artery ligation by resonant scattering of cytochromes,[22-24] which provides higher sensitivity and selectivity for the Raman signals as compared with the conventional Raman spectroscopy. The proper excitation wavelength, which is electronically resonant with the electronic transition of target molecules, allowed us to detect Raman spectra of cytochrome b5 and c in non-infarcted cardiac muscle of rat heart, having signal intensity $10^3$ - $10^5$ times stronger than that of the non-resonance signal.[9,25] In contrast to the rat experimental model, where infarcted tissues are distinctly replaced by fibrosis consisting mainly of type I collagen with no other major components,[26] accurate Raman spectroscopic evaluation is quite difficult in human old MI because of the spectral complexity of the tissue constituents and lack of relevant characteristic molecules exhibiting the resonant Raman scattering in the infarcted tissues. Analogously, the human myocardium may more or less present different Raman spectra between non-infarcted and infarcted myocardium; however, still unclear are the definitive Raman spectral fingerprints and spectroscopic criteria for evaluation of human old MI.

In this study, we sought to identify Raman spectral fingerprints and a prediction model for the evaluation of human old MI. To elucidate definitive Raman spectral fingerprints and a reliable prediction model for the evaluation of myocardial viability under the presence of noise in Raman spectrum, we employed a multivariate spectral analysis method, partial least squares regression-discriminant analysis (PLS-DA). PLS-DA employs a wide region of Raman spectra to construct a prediction model, enabling the reduction of noise effect that appears in the whole Raman spectrum on the prediction model.[23,27] Furthermore, the PLS-DA calculates latent variables (LVs) and scores, which gives us important spectral information for the prediction of tissue species.[23,27] We successfully obtained Raman spectra of non-infarcted and infarcted regions of human myocardium resected for volume reduction from five patients with chronic ischemic cardiomyopathy under left ventricular reconstruction (LVR). Our novel application of Raman spectroscopy to the human heart would be a useful means for the detection of myocardial viability during surgery.

## Results
**Raman spectra of the myocardial tissue samples excised surgically from human heart**

The Raman spectra of the non-infarcted and infarcted myocardium showed their own distinct features (Figure 1). In all the specimens excised from 5 patients, the non-infarcted myocardium, which predominantly consists of cardiomyocytes, exhibited four evident Raman peaks at 755, 1133, 1318, and 1590 $cm^{-1}$ that were all assigned to heme proteins[23] with no obvious differences among the patients (Figure 1b). In contrast, the Raman spectra of the infarcted myocardium were large in variation among the patients (Figure 1c). In addition, the infarcted myocardium had a spectral peak at 2942 $cm^{-1}$, which is distinct from the peak at 2934 $cm^{-1}$ in the non-infarcted myocardium. The spectral shapes at 1248, 1453, 1661, and 2942 $cm^{-1}$ were nearly identical to those of collagens (Supplemental Figure 1), a major constituent of the fibrotic tissue of old myocardial infarct, suggesting the presence of collagens in the infarcted myocardium as identified in our previous study on the rat heart.[23] Similar to the non-infarcted myocardium, the four evident peaks assigned to heme proteins at 755, 1133, 1318, and 1590 $cm^{-1}$ were also identified in the infarcted myocardium.

We compared peak intensities of the four representative Raman bands at 755, 1133, 1318, and 1590 $cm^{-1}$ of the non-infarcted and infarcted myocardium (Figure 2). The intensities of non-infarcted myocardium were significantly larger than those of the infarcted myocardium, indicating that the relative contribution of molecular compositions assigned to these bands was much larger in the non-infarcted myocardium than in the infarcted myocardium. Predominant chemical components assigned to these Raman bands might be different in these two regions, i.e., cytochrome c in the non-infarcted myocardium and heme proteins other than cytochrome c in the infarcted myocardium as described in the Discussion section. As a result, the Raman bands of cytochrome c were significantly emphasized in the non-infarcted myocardium, and the intensity variance of non-infarcted myocardium among individual patients might indicate the interpatient variability of reduced cytochrome c in residual cardiomyocytes, which is an important index of viability of the myocardium.

**Prediction of the myocardial tissue characteristics from single Raman bands**

We sought to predict tissue species by a simple approach using a single representative Raman band of non-infarcted and infarcted myocardium. We evaluated three representative Raman spectral features, i.e., peak intensity at 687 cm$^{-1}$, peak intensity at 755 cm$^{-1}$, and peak position ranging from 2930 to 2950 cm$^{-1}$.

For the Raman band at 755 cm$^{-1}$, prediction performance evaluated by using ROC analysis indicates that the sensitivities of non-infarcted and infarcted myocardium attained were 83.0% and 87.6%, respectively (Figure 3a). Although high prediction performance was obtained with the area under the curve (AUC) of 94.0% by using the peak intensity at 755 cm$^{-1}$, this Raman band reflected not only cytochrome c, but also the other heme proteins such as cytochrome b5, myoglobin, and hemoglobin[23].

To confirm the contribution of cytochrome c in the peak intensity at 755 cm$^{-1}$, we obtained the peak intensity at 687 cm$^{-1}$, which is one of the signatures of the presence of cytochrome c.[23] However, prediction performance of the peak intensity at 687 cm$^{-1}$ was quite low with the AUC of 62.8%, and the sensitivities of non-infarcted and infarcted myocardium were only 65.6% and 55.0%, respectively (Figure 3b). This indicates difficulty in confirming the presence of cytochrome c under the experimental conditions, i.e., 10-s exposure of 532-nm excitation at 235 to 300 μW/μm$^2$. Higher signal-to-noise ratio detection with longer exposure time or higher excitation laser power would improve the prediction performance; however, long exposure time might perturb the procedures of cardiac surgery, and higher excitation laser power might induce photo-induced damage to the specimen.

The peak position ranging from 2930 to 2950 cm$^{-1}$ is also the representative signature of the difference between non-infarcted and infarcted myocardium, indicating the presence of collagen in the infarcted myocardium. The ROC analysis of the peak position provided the AUC of 88.1%. The sensitivities of non-infarcted and infarcted myocardium were 91.8 % and 80.8 %, respectively.

These results indicated that the prediction of non-infarcted and infarcted myocardium was realized by using single Raman bands. The Raman bands representing the contribution of heme proteins (755 cm$^{-1}$) and collagen (peak position around 2930 to 2950 cm$^{-1}$) would be key signatures for the prediction of tissue species. However, prediction performance using a single band was limited in terms of signal-to-noise ratio of signal intensity.

**Reliable prediction for the discrimination of the infarcted myocardium from non-infarcted myocardium by PLS-DA**

To construct a more reliable prediction model for the evaluation of myocardial viability under the presence of noise in Raman spectrum, we employed a multivariate spectral analysis, PLS-DA. The PLS-DA provides LVs and scores that reflect spectral features of non-infarcted and infarcted myocardium. The LVs provide common spectral features of a Raman spectral data set employed in PLS-DA in terms of the difference of tissue species. The scores on individual LVs represent the contribution of the LVs on each observed Raman spectrum. The LVs thus have implications for assignment of Raman bands in terms of the statistical differences between non-infarcted and infarcted myocardium when combined with scores of PLS-DA. Furthermore, the distribution of score plot indicates a similarity of the spectral data set of each tissue species and patient in terms of LVs, showing predictability of non-infarcted and infarcted myocardium. LVs and scores of PLS-DA calculated with a total of 10,000 Raman spectra of non-infarcted and infarcted myocardium in five patients are shown in Figure 4.

Both scores of LV1 of the non-infarcted and infarcted myocardium of all patients were distributed in the negative value region (Figure 4a and Supplemental Movie 1). The scores of LV1 of the non-infarcted myocardium were smaller than those of the infarcted myocardium. The spectral peaks of LV1 included 1160, 1248, and 2942 cm$^{-1}$, which represented specific Raman bands of infarcted myocardium (Figure 4b). Although a few specific spectral features of non-infarcted myocardium, such as 687, 1177, and 1366 cm$^{-1}$, also appeared in LV1, the other specific spectral features of non-infarcted myocardium, such as 648, 2861, and 2934 cm$^{-1}$, were not included in LV1. In LV2, the Raman bands included 648, 687, 1177, 1366, 2861, and 2934 cm$^{-1}$, which were all specific in the non-infarcted myocardium, while no spectral feature of the infarcted myocardium appeared in LV2 (Figure 4c). The scores of LV2 of non-infarcted and infarcted myocardium were distributed in the positive region and negative region, respectively (Figure 4a and Supplemental Movie 1). As a result, LV1 and LV2 predominantly represented the important spectral features for the discrimination of non-infarcted and infarcted myocardium. Raman bands appearing in both the negative direction of LV1 and the positive direction of LV2 seemed to indicate the spectral features of non-infarcted myocardium. In contrast, Raman bands of LV1 in the negative direction with the subtraction of LV2 in the positive direction represented the spectral features of infarcted myocardium.

In LV3, a derivative-like spectrum was obtained, especially representative Raman bands of heme proteins, such as 755, 1133, 1318, and 1590 cm$^{-1}$ (Figure 4d). The scores of LV3 were roughly distributed depending on patients, not on tissue species (Figure 4a and Supplemental Movie 1). This result indicated that LV3 represented the difference among patients.

By using the prediction model derived by the PLS-DA, we sought to predict tissue species of non-infarcted and infarcted myocardium. Cross-validated detection accuracy of non-infarcted and infarcted myocardium is shown in Table 1. High cross-validated sensitivities and specificities of 99.98% and 99.92%, respectively, were obtained for the discrimination of non-infarcted and infarcted myocardium with the detection accuracy of 99.95%. This highly accurate detection capacity for tissue species supported the notion that our findings for the spectral features in the LVs of PLS-DA provided important information about differences of the Raman spectra of tissue species. Furthermore, these results also demonstrated the fundamental feasibility of the Raman spectroscopy for the evaluation of old MI of patients on the basis of the constituent molecules.

**Raman imaging of myocardium of marginal region of myocardial infarction**

Raman imaging of non-infarcted and infarcted myocardium was performed by using PLS-DA. For reconstruction of Raman images, we obtained PLS-DA-predicted values defined as the scalar product of regression vectors of each tissue species at each pixel. Representative Raman images of the marginal area of non-infarcted and infarcted myocardium obtained from patient 3 revealed that the PLS-DA-predicted values of non-infarcted and infarcted myocardium highlighted these tissue species, and visualized the boundary of non-infarcted and infarcted myocardium (Figure 5). These tissue distributions were consistent with the histology confirmed by HE- and Azan-stained serial section. Furthermore, fibrotic tissues that were present among cardiomyocytes were also visualized. Similar images were obtained for all the tissues from the other 4 patients (data not shown). These results indicated that the Raman spectroscopy with the PLS-DA has a capability to visualize non-infarcted and infarcted myocardium of human myocardium without staining.

**Discussion**

Our previous studies provided a proof-of-principle demonstration of Raman spectroscopy for the label-free evaluation of MI by using rat samples.[22,23] Here, we extended Raman spectroscopy to the human myocardium, and showed its feasibility for old MI by means of PLS-DA. We found that the Raman spectra of the human infarcted myocardium comprised various spectral components that did not appear in our previous rat study, while the spectral features of non-infarcted myocardium were almost identical to those in the rat.[22,23] Such varieties of Raman spectra of human infarcted myocardium hampered the discrimination of infarcted myocardium from non-infarcted myocardium. In the present study, our multivariate analytical method enables the prediction of non-infarcted and infarcted myocardium with high prediction accuracy.

In our previous study, we confirmed that strong Raman peaks at 755, 1133, and 1590 $cm^{-1}$ of rat heart were assigned to the specific vibrational modes originated from a porphyrin ring in the center of heme proteins. In addition to these bands, the Raman bands at 648, 687, and 1366 $cm^{-1}$ indicated the presence of reduced cytochrome c.[23] Thus, the Raman spectra obtained from human non-infarcted myocardium that included these Raman bands also seemed to have arisen from reduced forms of cytochromes c in cardiomyocytes (Figure 1a). Reduced cytochrome c is generally present in viable cardiomyocytes. In contrast, there were no Raman bands at 648 and 687 $cm^{-1}$ in the infarcted myocardium, indicating the Raman bands at 755, 1133, and 1590 cm$^{-1}$ were assigned to heme proteins other than cytochrome c, such as hemoglobin, myoglobin, and other cytochromes.[23] Heme proteins are not common molecules in infarcted myocardium, which might appear by blood deposition during surgery or other contaminations following repair process of human old MI. As a result, patient variability of Raman spectra appeared in infarcted myocardium, but barely appeared in non-infarcted myocardium.

The intensities of the Raman bands at 755, 1133, 1318, and 1590 cm$^{-1}$ in non-infarcted myocardium tend to depend on the intensities of these Raman bands of non-infarcted myocardium (Figure 1). Especially in patient 4, large contribution of these Raman bands was obtained in the infarcted myocardium, and the intensities of these Raman bands in the non-infarcted myocardium were relatively larger than that in the other patients. This result indicated that the contributions of heme proteins appearing in the infarcted myocardium seemed to also appear in the non-infarcted myocardium; in other words, homogeneous background of the Raman spectrum of heme proteins might be present. The Raman spectra of non-infarcted and infarcted myocardium of each patient were obtained at near borders of MI, in which the distances between these observation points in non-infarcted and infarcted myocardium were within a few millimeters. One possible case for these homogeneous contributions of heme proteins in the non-infarcted and infarcted myocardium is the deposition of hemoglobin of blood. The myocardium was excised following LVR, thus blood must be exposed in the myocardium by hemorrhage following cardiac surgery. No red blood cells were observed in terms of HE-stained sections, indicating that the deposition of hemoglobin of hemolyzed blood might contribute to the Raman spectra in some cases.

In Raman spectroscopy, visible or near-infrared excitation light is irradiated to the specimen and observed inelastic scattering light reflects molecular information about the myocardium via Raman spectrum. Since contrast agents and any preparation are not needed, Raman spectroscopy has potential for intraoperative observation with minimal invasiveness. Indeed, several groups initiated clinical trials of Raman spectroscopy for label-free *in vivo* diagnosis in the fields of breast surgery,[28] gastrointestinal surgery,[29] and skin surgery.[30] Although further confirmation of the Raman spectroscopic technique under *in vivo* condition is required for the application to cardiac surgery, we anticipate that label-free discrimination of non-infarcted and infarcted myocardium will allow us intraoperative evaluation of myocardial viability of human patients in the future.

## **Conclusions**

This is the first report demonstrating the evaluation of old MI in patients undergoing cardiac surgery by means of Raman spectroscopy employing PLS-DA. Our results revealed the feasibility of the Raman spectroscopy for the label-free assessment of myocardial viability of human myocardium.

Our proposed approach may be a promising technique for direct evaluation of myocardial viability during surgery in the future.

## Methods

**Patients and sample preparation**

All experiments were performed with the approval of the Ethics Committee of Kyoto Prefectural University of Medicine (Permission No. RBMR-C-966). For all patients, fully informed consent was obtained after a full explanation of the study design and before the surgery. Myocardium was resected from five patients with old myocardial infarct under LVR at University Hospital, Kyoto Prefectural University of Medicine, conducted between August 2012 and January 2014 during cardiac surgery. The excised myocardium kept cold on ice was quickly transferred to the laboratory, embedded in frozen section compound (FSC 22; Leica, Wetzlar, Germany), snap-frozen in a freezing mixture (frozen carbon dioxide with acetone), and stored at -80°C until experiment. The frozen samples were sliced into 5-μm-thick sections with a cryostat microtome (CM1900; Leica). Three serial sections were obtained from each tissue for Raman and histological analysis. The sections for the Raman analysis were mounted on a 0.17-mm-thick cover glass without fixation. The sections for the histological analysis were subjected to hematoxylin and eosin (HE) staining or Azan staining to confirm histology.

**Raman spectroscopy**

Raman spectra and the reflecting spectral images were acquired with a slit-scanning confocal Raman microscope (RAMAN-11; Nanophoton, Osaka, Japan) as described previously.[19,22,23,31] A frequency doubled Nd:YAG laser (532 nm) was employed for excitation. The excitation laser beam was focused into a line on a sample through a cylindrical lens and a high-magnitude water-immersion objective lens (UPLSAPO x60, NA = 1.2; Olympus, Tokyo, Japan) or a low-magnitude objective lens (UPlanFL N, x10, NA = 0.3; Olympus, Tokyo, Japan). The high-magnitude water-immersion objective lens was used for the acquisition of Raman spectra. These Raman spectra were also used for the construction of a prediction model with PLS-DA analysis in two-dimensional imaging. The low-magnitude objective lens was used for the acquisition of two-dimensional images.

Raman scattered light was collected with the same objective lens. The Raman scattered light was focused onto the input slit with 70 μm width of the spectrometer with a 600 grooves/mm grating, and detected with a two-dimensional image sensor (Pixis 400BR, -70°C, 1340x400 pixels; Princeton Instruments, Trenton, NJ, USA). Owing to the line shaped focus of the excitation beam, a one-dimensional spectral image (one-dimensional space and Raman spectrum) from the line-illuminated site was simultaneously obtained. Each pixel data in space was used for a single data point in all spectral analysis. Two-dimensional Raman spectral images were obtained by scanning the line-shaped laser focus in a single direction. The excitation laser power and the exposure time was set at from 235 to 300 $\mu W/\mu m^2$ on the sample plane, and 10 s, respectively.

**Preprocessing of spectral analysis**

Before spectral analysis, all raw Raman spectra were calibrated using the known lines of ethanol. The broad fluorescence background was fitted with the modified least-squares polynomial curve[32] and subtracted it from calibrated Raman spectra with Nanophoton software (Nanophoton, Osaka, Japan). The background-removed Raman spectra were normalized with the integrated intensity of each Raman spectrum.

**Statistical analysis of prediction capacity by means of a single band Raman spectrum**

The prediction capacity by means of a single band Raman spectrum was evaluated by using the peak intensity of 687 $cm^{-1}$, peak intensity of 755 $cm^{-1}$, and peak position of 2930 to 2950 $cm^{-1}$. The peak intensities of 687 and 755 $cm^{-1}$ were obtained in terms of area intensity as shown in Supplementary Figure 2 to extract proper intensity of each Raman bands. To extract peak position of 2930 to 2950 $cm^{-1}$ against noise contribution, Raman spectrum around 2940 $cm^{-1}$ was fitted by Gaussian function in IGOR Pro software (WaveMetrics, OR, USA). Peak position comprised in the fitted Gaussian function was obtained for the statistical analysis. These spectral analyses were performed with programs written in programing language in MATLAB (Mathworks, Natick, MA, USA) and IGOR Pro software (WaveMetrics, OR, USA).

Receiver operating characteristic (ROC) analysis was employed to predict tissue species by using single band of Raman spectrum. ROC analysis was performed with programs written in the IGOR programing language in IGOR Pro software (WaveMetrics, OR, USA). Optimal sensitivities of non-infarcted and infarcted myocardium were defined as those yielding the minimal value for $(1-\text{sensitivity}_{\text{non-infarcted}})^2 + (1-\text{sensitivity}_{\text{infarcted}})^2$, where $\text{sensitivity}_{\text{non-infarcted}}$ and $\text{sensitivity}_{\text{infarcted}}$ respectively indicated the sensitivities of non-infarcted and infarcted myocardium.

**PLS-DA analysis for the prediction of tissue species**

PLS-DA analysis was employed for the construction of a prediction model of old MI. PLS-DA analysis is widely applied to construct quantitative prediction model for classification in spectral analysis by projecting the predicted and observable variables into a new space to maximize the covariance between the response and independent variables. All calculations in PLS-DA were performed with PLS toolbox (eigenvector Research, WA, USA) in MATLAB (Mathworks, Natick, MA, USA).

Randomly obtained 1000 Raman spectra in each tissue species of each patient were analyzed. Since the cover glass retaining frozen sections exhibited Raman spectra from around 850 to 1020 cm$^{-1}$, and Raman-silent region from around 2000 to 2800 cm$^{-1}$ exhibited no significant Raman spectra in myocardium, we extracted Raman spectra from 596 to 836 cm$^{-1}$, from 1024 to 1723 cm$^{-1}$, and from 2818 to 2980 cm$^{-1}$ of each individual spectrum for PLS-DA analysis. We constructed a prediction model with orthogonal condition and calculated LVs and scores. Leave-one-out cross validation analysis was employed for the verification of the predictive accuracy of the prediction model using PLS-DA[33-35]. The optimal number of LVs used in the prediction model was determined as first minimum value of number of misclassification rate as shown in Supplementary Figure 3.

For reconstruction of two-dimensional Raman images, we obtained the PLS-DA-predicted values defined as the scalar product of a regression vector with the Raman spectrum at each pixel. Original Raman spectral-images were obtained with 400x100 pixels in spatial dimension, and were binning

The regression vector was derived by PLS-DA analysis in the construction process of a prediction model as shown above. The Raman spectra of 596 to 836 cm$^{-1}$, from 1024 to 1723 cm$^{-1}$, and from 2818 to 2980 cm$^{-1}$ were used for the reconstruction of two-dimensional prediction map of tissue species.

**Raman spectroscopy of pure chemical of collagen**

Powdered collagen type I derived from bovine Achilles tendon (C9879, Sigma-Aldrich, MO, USA) was placed on a slide glass, and observed a Raman spectrum using the slit-scanning confocal Raman microscope as shown above.

**References**

1. Klein, C. *et al.* Assessment of myocardial viability with contrast-enhanced magnetic resonance imaging: comparison with positron emission tomography. *Circulation* **105**, 162-167, (2002).

2. Ogawa, M. *et al.* Surgical ventricular restoration based on evaluation of myocardial viability with delayed-enhanced magnetic resonance imaging. *Gen. Thorac. Cardiovasc. Surg.* **55**, 149-157; discussion 157, (2007).

3. Knapp, F. F., Jr., Franken, P. & Kropp, J. Cardiac SPECT with iodine-123-labeled fatty acids: evaluation of myocardial viability with BMIPP. *J. Nucl. Med.* **36**, 1022-1030, (1995).

4. Wu, K. C. & Lima, J. A. Noninvasive imaging of myocardial viability: current techniques and

future developments. *Circ. Res.* **93**, 1146-1158, (2003).

5       Bax, J. J., van der Wall, E. E. & Harbinson, M. Radionuclide techniques for the assessment of myocardial viability and hibernation. *Heart* **90 Suppl 5**, v26-33, (2004).

6       Underwood, S. R. *et al.* Imaging techniques for the assessment of myocardial hibernation. Report of a Study Group of the European Society of Cardiology. *Eur. Heart J.* **25**, 815-836, (2004).

7       Puppels, G. J. *et al.* Studying single living cells and chromosomes by confocal Raman microspectroscopy. *Nature* **347**, 301-303, (1990).

8       van Manen, H. J., Kraan, Y. M., Roos, D. & Otto, C. Single-cell Raman and fluorescence microscopy reveal the association of lipid bodies with phagosomes in leukocytes. *Proc. Natl. Acad. Sci. U. S. A.* **102**, 10159-10164, (2005).

9       Okada, M. *et al.* Label-free Raman observation of cytochrome c dynamics during apoptosis. *Proc. Natl. Acad. Sci. U. S. A.* **109**, 28-32, (2012).

10      Hattori, Y. *et al.* In vivo Raman study of the living rat esophagus and stomach using a micro-Raman probe under an endoscope. *Appl. Spectrosc.* **61**, 579-584, (2007).

11      Freudiger, C. W. *et al.* Label-free biomedical imaging with high sensitivity by stimulated Raman scattering microscopy. *Science* **322**, 1857-1861, (2008).

12      Ozeki, Y. *et al.* High-speed molecular spectral imaging of tissue with stimulated Raman scattering. *Nat. Photonics* **6**, 844-850, (2012).

13      Kim, S. H. *et al.* Multiplex coherent anti-Stokes Raman spectroscopy images intact atheromatous lesions and concomitantly identifies distinct chemical profiles of atherosclerotic lipids. *Circ. Res.* **106**, 1332-1341, (2010).

14      Mizuno, A., Kitajima, H., Kawauchi, K., Muraishi, S. & Ozaki, Y. Near‐infrared Fourier transform Raman spectroscopic study of human brain tissues and tumours. *J. Raman Spectrosc.* **25**, 25-29, (1994).

15      Brennan, J. F., 3rd *et al.* Determination of human coronary artery composition by Raman spectroscopy. *Circulation* **96**, 99-105, (1997).

16      Caspers, P. J., Lucassen, G. W. & Puppels, G. J. Combined in vivo confocal Raman spectroscopy and confocal microscopy of human skin. *Biophys. J.* **85**, 572-580, (2003).

17      Redd, D. C., Feng, Z. C., Yue, K. T. & Gansler, T. S. Raman spectroscopic characterization of human breast tissues: implications for breast cancer diagnosis. *Appl. Spectrosc.* **47**, 787-791, (1993).

18      Frank, C. J., McCreery, R. L. & Redd, D. C. Raman spectroscopy of normal and diseased human breast tissues. *Anal. Chem.* **67**, 777-783, (1995).

19      Minamikawa, T. *et al.* Label-free detection of peripheral nerve tissues against adjacent tissues by spontaneous Raman microspectroscopy. *Histochem. Cell Biol.* **139**, 181-193, (2013).

20    Minamikawa, T., Harada, Y. & Takamatsu, T. Ex vivo peripheral nerve detection of rats by spontaneous Raman spectroscopy. *Sci. Rep.* **5**, 17165, (2015).

21    Kumamoto, Y., Harada, Y., Tanaka, H. & Takamatsu, T. Rapid and accurate peripheral nerve imaging by multipoint Raman spectroscopy. *Sci. Rep.* **7**, 845, (2017).

22    Ogawa, M. *et al.* Label-free biochemical imaging of heart tissue with high-speed spontaneous Raman microscopy. *Biochem. Biophys. Res. Commun.* **382**, 370-374, (2009).

23    Nishiki-Muranishi, N. *et al.* Label-free evaluation of myocardial infarction and its repair by spontaneous Raman spectroscopy. *Anal. Chem.* **86**, 6903-6910, (2014).

24    Ohira, S. *et al.* Label-free detection of myocardial ischaemia in the perfused rat heart by spontaneous Raman spectroscopy. *Sci. Rep.* **7**, 42401, (2017).

25    Hamada, K. *et al.* Raman microscopy for dynamic molecular imaging of living cells. *J. Biomed. Opt.* **13**, 044027, (2008).

26    Kumar, V., Abbas, A. K., Fausto, N., Robbins, S. L. & Cotran, R. S. *Robbins and Cotran Pathologic basis of disease*. 7th edn,  (Elsevier Saunders, 2005).

27    Bergholt, M. S. *et al.* Raman endoscopy for in vivo differentiation between benign and malignant ulcers in the stomach. *Analyst* **135**, 3162-3168, (2010).

28    Haka, A. S. *et al.* In vivo margin assessment during partial mastectomy breast surgery using raman spectroscopy. *Cancer Res.* **66**, 3317-3322, (2006).

29    Shim, M. G., Song, L. M., Marcon, N. E. & Wilson, B. C. In vivo near-infrared Raman spectroscopy: demonstration of feasibility during clinical gastrointestinal endoscopy. *Photochem. Photobiol.* **72**, 146-150, (2000).

30    Caspers, P. J., Lucassen, G. W., Wolthuis, R., Bruining, H. A. & Puppels, G. J. In vitro and in vivo Raman spectroscopy of human skin. *Biospectroscopy* **4**, S31-39, (1998).

31    Harada, Y. *et al.* Intracellular dynamics of topoisomerase I inhibitor, CPT-11, by slit-scanning confocal Raman microscopy. *Histochem. Cell Biol.* **132**, 39-46, (2009).

32    Lieber, C. A. & Mahadevan-Jansen, A. Automated Method for Subtraction of Fluorescence from Biological Raman Spectra. *Appl. Spectrosc.* **57**, 1363-1367, (2003).

33    Magee, N. D. *et al.* Ex vivo diagnosis of lung cancer using a Raman miniprobe. *J. Phys. Chem. B* **113**, 8137-8141, (2009).

34    Tan, Y., Shi, L., Tong, W., Gene Hwang, G. T. & Wang, C. Multi-class tumor classification by discriminant partial least squares using microarray gene expression data and assessment of classification models. *Comput. Biol. Chem.* **28**, 235-243, (2004).

35    Bergholt, M. *et al.* In vivo diagnosis of esophageal cancer using image-guided Raman endoscopy and biomolecular modeling. *Technol. Cancer Res. Treat.* **10**, 103-112, (2011).

**Acknowledgments**

A portion of this work was supported by the Adaptable and Seamless Technology Transfer Program through Target-driven R&D (AS2321506F) from the Japan Science and Technology Agency (JST), JSPS KAKENHI (JP11J08134, JP16K16389) from the Japan Society for the Promotion of Science (JSPS), and PRESTO (JPMJPR17PC) from JST. We thank Dr. Mitsugu Ogawa of John Hunter Hospital, Australia, for his helpful discussion. We also acknowledge Mr. Toshifumi Kawamura and Mr. Takashi Okuda of the Department of Pathology and Cell Regulation, Kyoto Prefectural University of Medicine, Japan, for histological staining.

**Author contributions**

T.Y., T.M., and T.T. conceived the project. T.Y. and T.M. designed and performed the experiments. T.Y. and H.Y. contributed to the sample preparation. T.Y. and T.M. analyzed the data. T.Y., T.M., Y.H., H.T. and T.T. contributed to manuscript preparation. All authors discussed the results and commented on the manuscript.

**Disclosures**

The authors declare competing financial interests. T.M., Y.H., and T.T. have filed a patent related to this work. T.T. is a technical adviser of Nanophoton Corp., Osaka, Japan.

**Materials and Correspondence**

Correspondence and requests for materials should be addressed to Takeo Minamikawa or Tetsuro Takamatsu.

Table 1. Detection power of non-infarcted and infarcted myocardium. Detection accuracy for the prediction of non-infarcted and infarcted myocardium was 99.95%.

|  | Histology | |
| --- | --- | --- |
|  | Non-infarcted | Infarcted |
| Non-infarcted | 4999 | 4 |
| Infarcted | 1 | 4996 |
| Sensitivity | 99.98% | 99.92% |

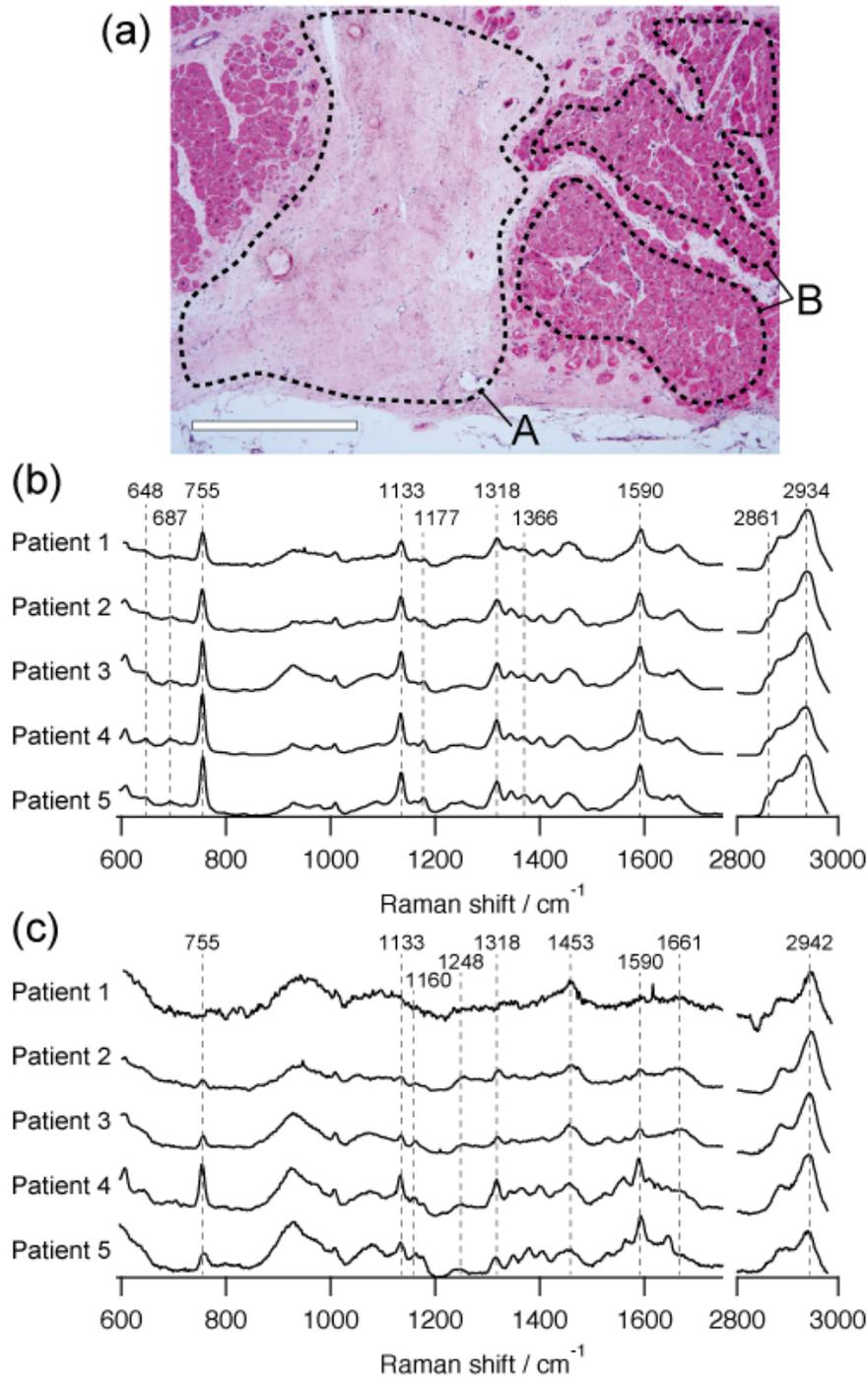

Figure 1. Raman spectra of the border zone myocardium obtained from 5 patients. (a) A representative HE-stained image of the boundary between the infarcted region (dashed area A) and non-infarcted region (dashed area B) of the heart tissue excised from patient 3. Scale bar represents 1 mm. Representative Raman spectra of non-infarcted myocardium (b) and infarcted myocardium (c) of each patient. Each spectrum was normalized by the integrated intensity of each Raman spectrum.

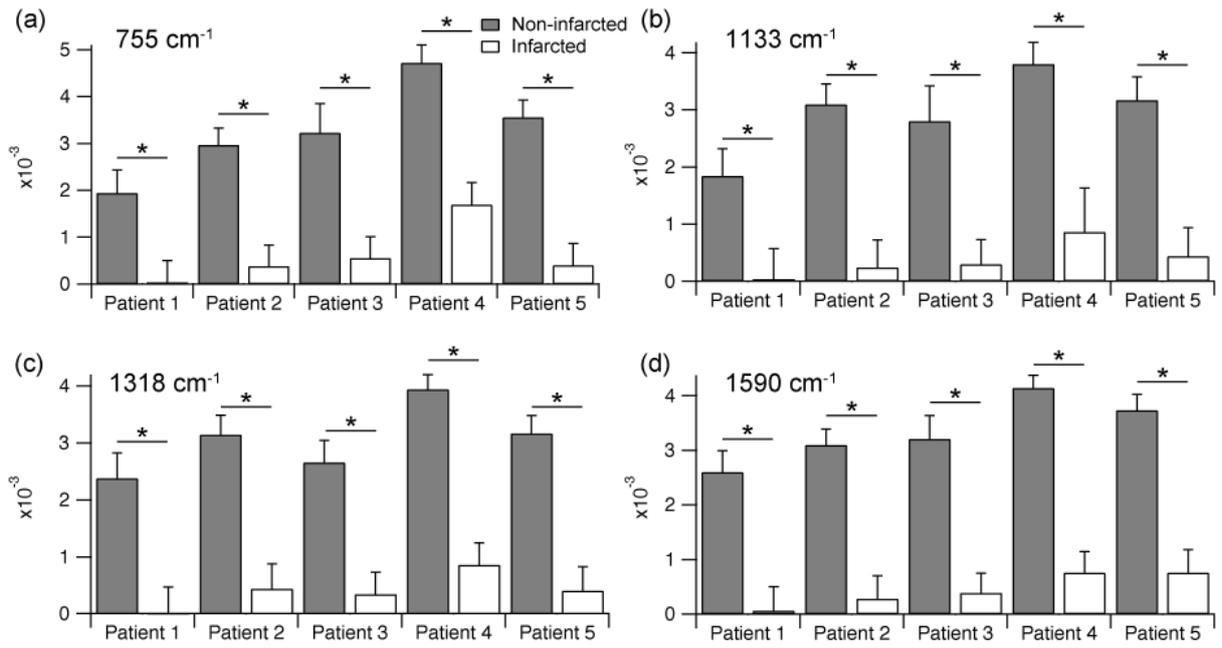

Figure 2. Peak intensities of representative Raman bands of the non-infarcted and infarcted myocardium obtained from 5 patients at 755 cm$^{-1}$ (a), 1133 cm$^{-1}$ (b), 1318 cm$^{-1}$ (c), and 1590 cm$^{-1}$ (d). Asterisks indicate significant differences ($p < 0.01$) by Student's t test.

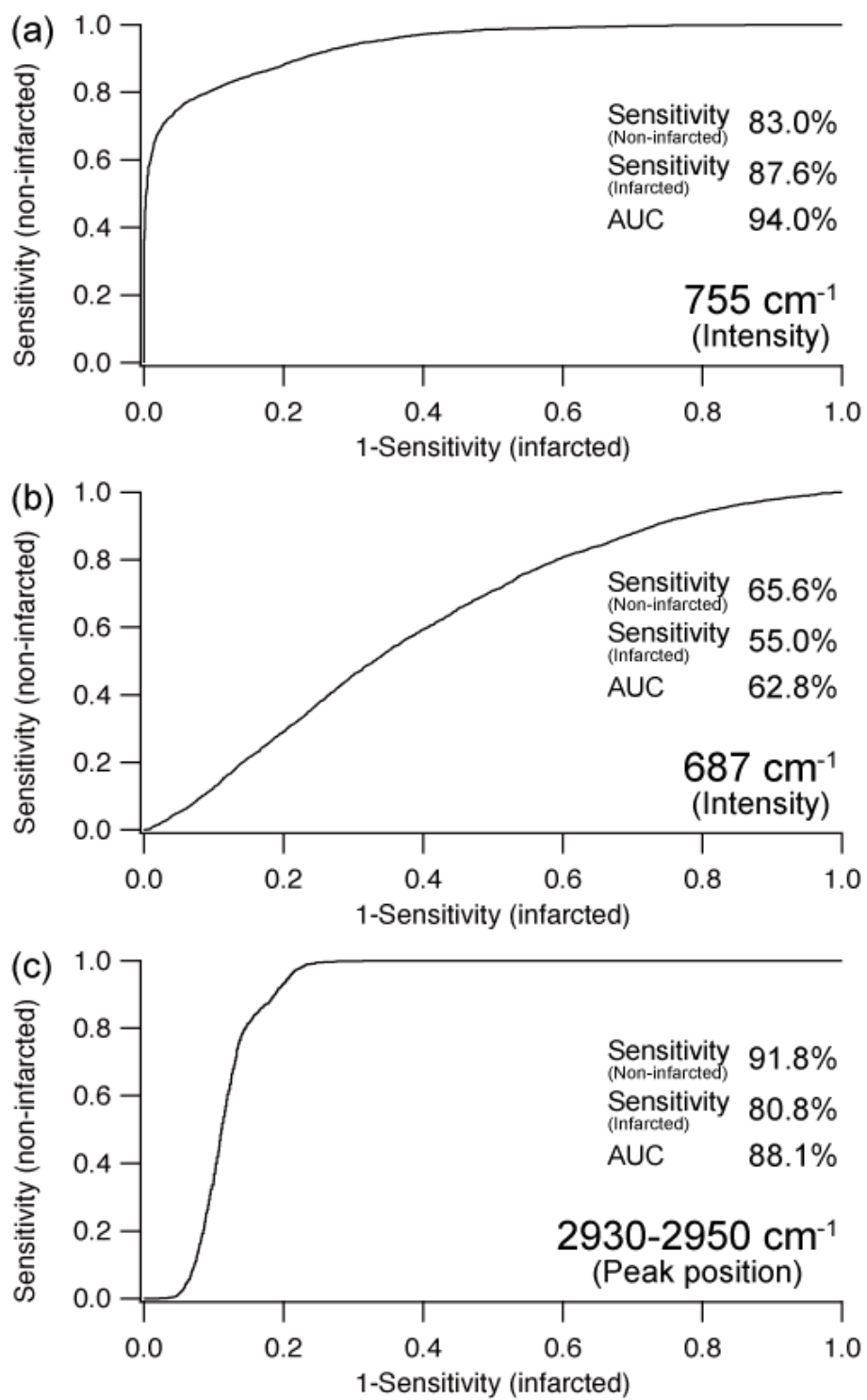

Figure 3. ROC analysis for the prediction of non-infarcted and infarcted myocardium of 5 patients by using representative Raman bands. Peak intensity of 755 cm$^{-1}$ (a), peak intensity of 687 cm$^{-1}$ (b), and peak position around 2930-2950 cm$^{-1}$ (c) were used for the ROC curve calculation.

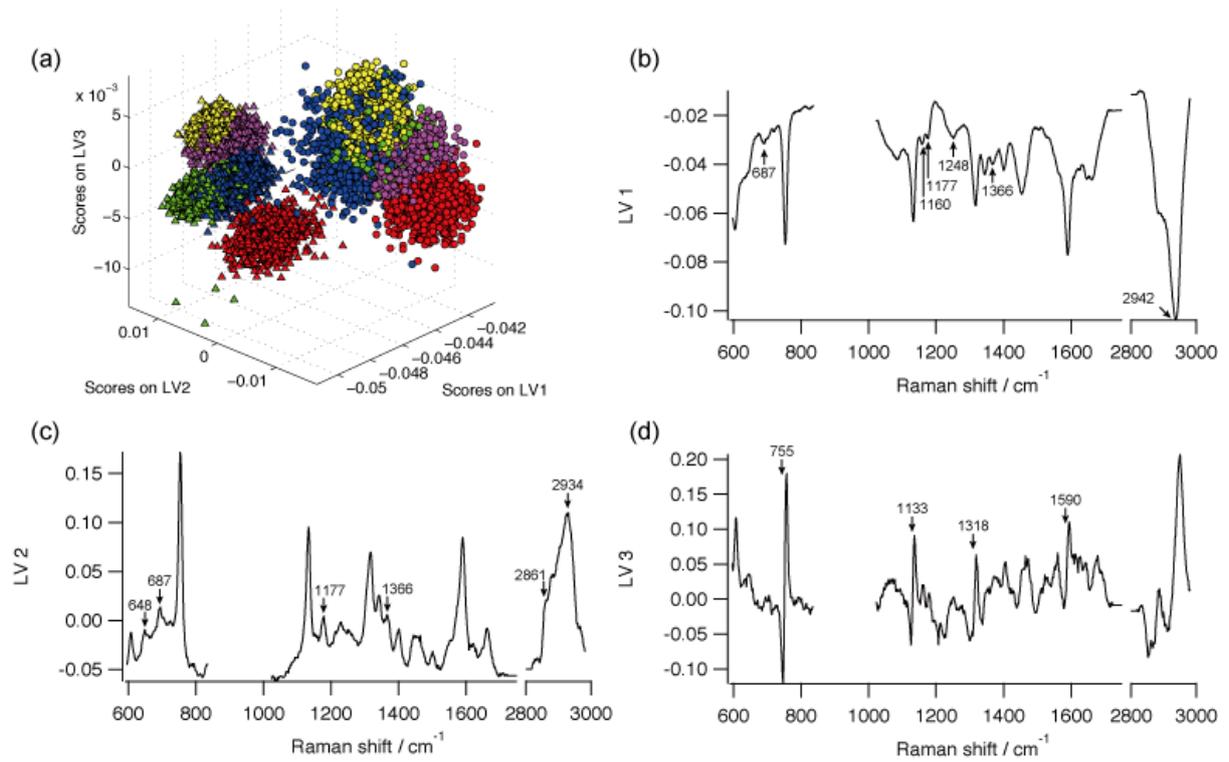

Figure 4. PLS-DA of non-infarcted and infarcted myocardium. (a) Score plot of LV1, LV2, and LV3. Triangle, non-infarcted cardiac tissue; circle, infarcted cardiac tissue; red, patient 1; green, patient 2; blue, patient 3; yellow, patient 4; and purple, patient 5. Contributions of LV1 (b), LV2 (c), and LV3 (d) to Raman spectra are 78.65%, 2.56%, and 0.41%, respectively.

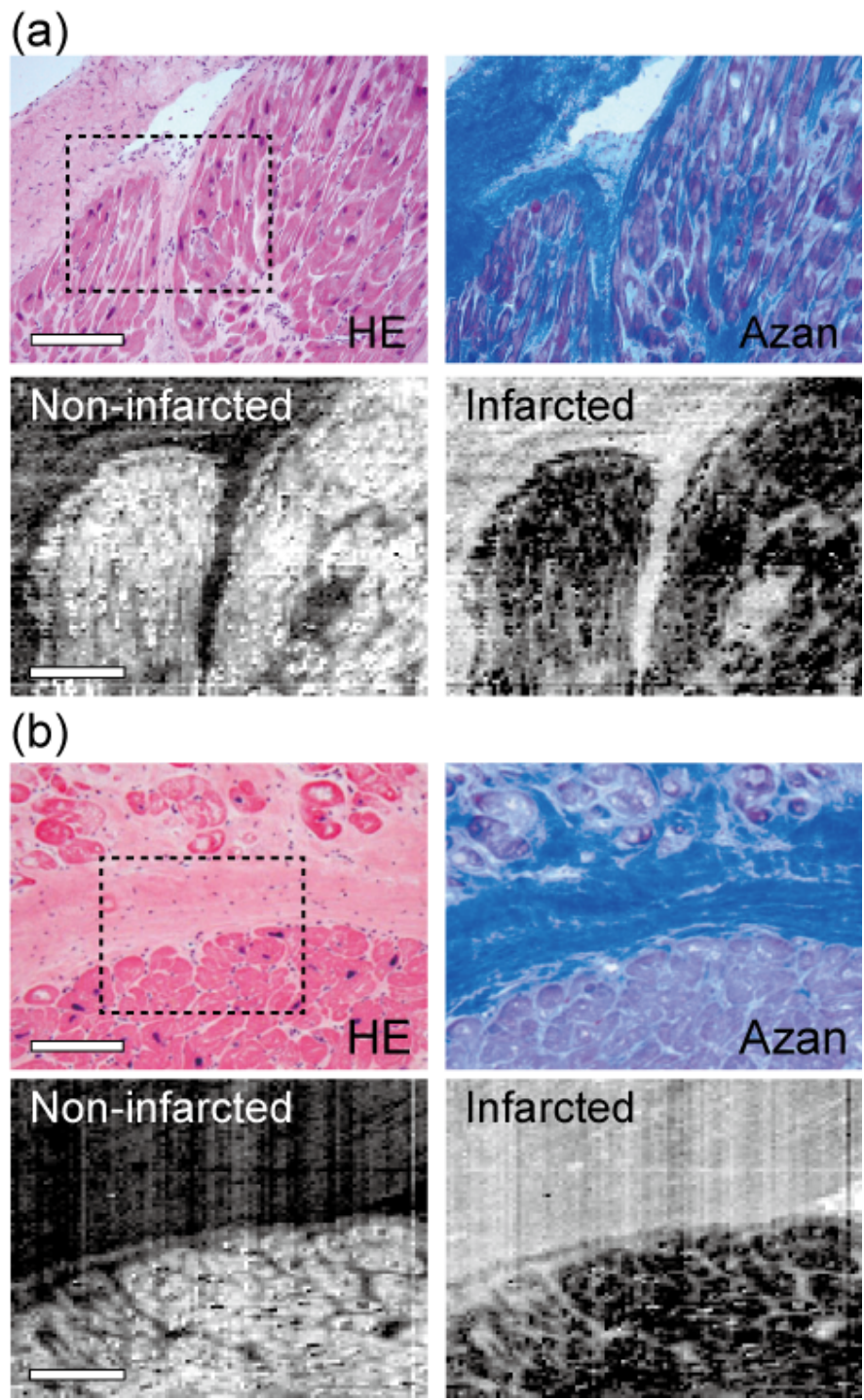

Figure 5. Representative 2D Raman images of a marginal region of non-infarcted and infarcted myocardium. (a, b) Two representative marginal regions of non-infarcted and infarcted myocardium of patient 3. Raman images at the dashed squares indicated in the HE-stained images were reconstructed by using PLS-DA-based prediction of tissue species. Scale bars in the HE-stained and Raman images in (a) and (b) are 200 μm and 100 μm, respectively.